# RECONSTRUCTION OF QUANTITATIVE SUSCEPTIBILITY MAPS FROM PHASE OF SUSCEPTIBILITY WEIGHTED IMAGING WITH CROSS-CONNECTED Ψ-NET

*Zhiyang Lu [a], Jun Li [b, c], Zheng Li [a], Hongjian He [b], Jun Shi [a]\**

[a] School of Communication and Information Engineering, Shanghai University, China
[b] Center for Brain Imaging Science and Technology, Zhejiang University, China
[c] Research Center for Healthcare Data Science, Zhejiang Lab, China

## ABSTRACT

Quantitative Susceptibility Mapping (QSM) is a new phase-based technique for quantifying magnetic susceptibility. The existing QSM reconstruction methods generally require complicated pre-processing on high-quality phase data. In this work, we propose to explore a new value of the high-pass filtered phase data generated in susceptibility weighted imaging (SWI), and develop an end-to-end Cross-connected Ψ-Net (CΨ-Net) to reconstruct QSM directly from these phase data in SWI without additional pre-processing. CΨ-Net adds an intermediate branch in the classical U-Net to form a Ψ-like structure. The specially designed dilated interaction block is embedded in each level of this branch to enlarge the receptive fields for capturing more susceptibility information from a wider spatial range of phase images. Moreover, the crossed connections are utilized between branches to implement a multi-resolution feature fusion scheme, which helps CΨ-Net capture rich contextual information for accurate reconstruction. The experimental results on a human dataset show that CΨ-Net achieves superior performance in our task over other QSM reconstruction algorithms.

## 1. INTRODUCTION

Quantitative susceptibility mapping (QSM) is a novel post-processing technique of magnetic resonance imaging (MRI) for quantifying magnetic susceptibility of tissues [1, 2]. It has great potential for diagnosis of various neurological disorders by revealing disease-related changes in myelin, iron and calcium deposition, etc. [3, 4]. In order to derive reliable susceptibility measures, the conventional approaches for QSM reconstruction generally require the high-quality raw phase data collected with multi-echo gradient echo (mGRE) sequences for instance, and also complicated processing steps including phase unwrapping, background phase removal and dipole inversion [5, 6]. Thus, it is a time-consuming imaging procedure that needs high computational complexity.

Recently, convolutional neural network (CNN), as a typical method of deep learning, has achieved great success in MRI reconstruction [7, 8, 9], which is also preliminarily used for QSM reconstruction [10, 11, 12]. These pioneering works mainly adopt U-Net as backbone to effectively learn the nonlinear mapping from the raw phase to QSM images [13]. However, these approaches still require phase data of sufficient quality. Their resultant maps need further improvement as well.

On the other hand, susceptibility weighted imaging (SWI) is a more widely used MRI technique in clinical practice, which combines both the magnitude and phase data from GRE sequences to enhance the susceptibility contrast [14, 15]. SWI generates plenty of high-pass filtered (HP-filtered) phase data during its imaging procedure under default clinical settings. Since the high-pass filtering causes a great loss on components about tissue susceptibility [16], the conventional methods cannot derive quantitative susceptibility maps from these low-quality HP-filtered phase data.

In fact, due to the powerful ability of CNN on learning nonlinear mapping, it has the potential to train a CNN model for reconstruction of QSM from these SWI phase data in an end-to-end way free from complicated pre-processing. To the best of our knowledge, there is no work about this approach.

It is a straightforward way to apply the existing U-Net-based QSM reconstruction algorithms to this new task. However, due to the great loss of components related to susceptibility in the HP-filtered phase data, these algorithms with coarse structures cannot learn sufficient information from the SWI phase for accurate QSM reconstruction. In particular, the susceptibility at a voxel depends on the phase information from the surrounding areas, and therefore, it is necessary to capture more susceptibility information from a wider spatial range of phase images. Unfortunately, these U-Net-based networks generally use small kernels in convolutions, leading to limited receptive fields. Besides, a U-Net model can hierarchically generate features with different resolutions, among which the high-resolution features reflect the local statistics, while the low-resolution features capture long-ranged dependencies in phase images. However, the skip connections designed in these U-Net-based reconstruction algorithms only propagate features of single resolution and ignore the contextual information from hierarchical features.

To this end, we propose a novel Cross-connected Ψ-Net (CΨ-Net) to directly reconstruct QSM from the HP-filtered phase data of SWI, in which an intermediate branch with several novel dilated interaction blocks (DIB) is added between the original two branches of U-Net to form a Ψ-like structure. Moreover, the crossed connections are designed between branches to aggregate contextual multi-resolution features. The main contribution is three-fold as follows:

1) A CΨ-Net is proposed to reconstruct susceptibility maps directly from the HP-filtered phase data of SWI. It explores a new value of the low-quality SWI phase and offers an efficient way to reconstruct QSM free from complicated pre-processing on phase.
2) A novel DIB is developed to enlarge the receptive fields and then learn more long-ranged spatial information about tissue susceptibility.
3) A multi-resolution feature fusion scheme implemented by crossed connections is designed to fuse multi-resolution features for effectively learning contextual information in CΨ-Net.

## 2. METHOD

As shown in Fig. 1, the CΨ-Net is developed based on a 3D U-Net architecture by adding an intermediate branch between the encoder and decoder branches to enlarge receptive fields with several DIBs. A multi-resolution feature fusion scheme is designed to aggregate features of different resolutions through the crossed connections between branches. The whole network takes the SWI HP-filtered phase images as input, and then reconstructs QSM under the guidance of a mean squared error (MSE) objective function.

### 2.1. Dilated Interaction Block (DIB)

The larger receptive fields can help capture more susceptibility information from far-distance regions in SWI HP-filtered phase images. Therefore, three consecutive DIBs are developed from bottom to top in the intermediate branch between the encoder and decoder branches to refine features and get robust representations with enlarged receptive fields.

Dilation convolution provides an efficient way to exploit information from a larger receptive field [17]. Therefore, as shown in Fig.1, a DIB consists of two convolutional layers, and each layer contains three paralleled 3×3×3 dilated convolution with the dilation rates of 1, 2 and 3, respectively, to generate features from different spatial regions. The dilation rate denotes the gap among weights in a kernel, and a dilated convolution equals to a standard convolution when the dilation rate is 1. In order to make full use of the paralleled features, the internal crossed connections are applied between two layers for feature propagation. That is, all the features generated from different dilated convolution in the first layer are jointly fed into each dilated convolution in the second layer. Finally, the three representations yielded from the second layer are concatenated as the output of the DIB.

Define the input feature map of the DIB to be $x$, the convolutional implementations in this block can be formulated as:

$$G(x) = [f^1_{r=1}(x), f^1_{r=2}(x), f^1_{r=3}(x)] \quad (1)$$

$$H(x) = [f^2_{r=1}(G(x)), f^2_{r=2}(G(x)), f^2_{r=3}(G(x))] \quad (2)$$

where $[\cdot]$ denotes the concatenation, $G(\cdot)$ and $H(\cdot)$ are the mappings of the first layer and the whole DIB, respectively, and $f^j_{r=i}$ is the function of the dilated convolution with dilation rate of $i$ at the $j^{th}$ layer. It is worth noting that more convolutional layers with the same structure can be cascaded with crossed connection in a DIB.

### 2.2. Multi-Resolution Feature Fusion

To fully utilize the contextual information in the hierarchical features of different resolutions, a multi-resolution feature fusion scheme is designed between branches in CΨ-Net, which is implemented by crossed connections.

This fusion scheme is firstly applied to merge the hierarchical features generated from the encoder branch for further refinement in DIBs. Specifically, as shown in Fig. 1, the input to a DIB contains information of three feature representations from different levels. Thus, the crossed connections are utilized to resample hierarchical features and then feed them into the corresponding DIBs. There are three resampling strategies in crossed connections. That is, the top-down connections downsample the features by convolution with stride 2 or 4, the down-top connections upsample the features by deconvolution with stride 2 or 4, and the parallel skip connections do not change the resolutions. Therefore, the set of features generated from each level of the encoder branch is resampled to generate three feature representations with different resolutions to match three DIBs, respectively. Consequently, a DIB receives three sets of resampled features from the encoder branch together with a set of upsampled features from the DIB in deeper level. All these features are then concatenated as the input of this DIB.

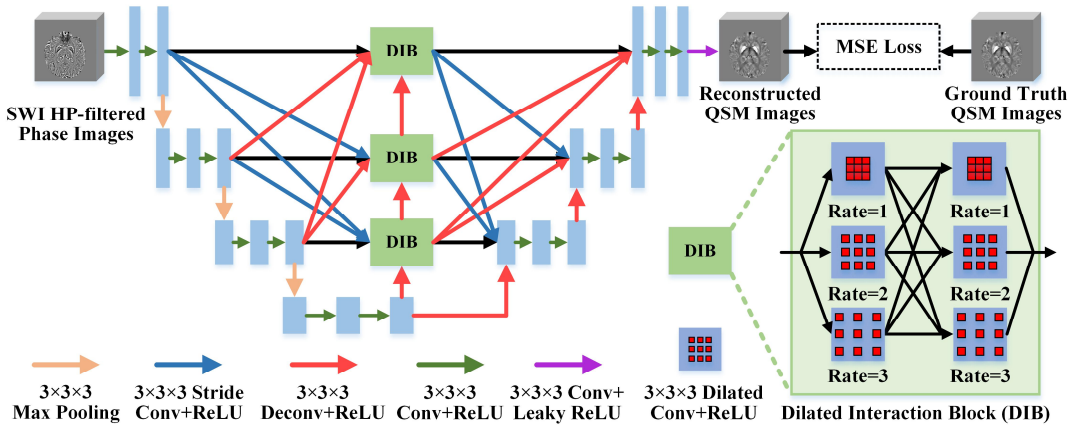

**Fig. 1.** The architecture of CΨ-Net.

The same implement is applied to conduct the fusion scheme between the DIBs and the decoder branch. Thus, we can capture rich contextual information of different resolutions for further QSM reconstruction.

### 2.3. QSM Reconstruction

The learned features from decoder branch are then fed to a convolutional layer with leaky ReLU to generate the reconstructed QSM images.

An MSE loss function is used to guide the reconstructed susceptibility maps to approximate the ground truth. Given a set of training date $\{I_{SWI}^i, I_{QSM}^i\}_{i=1}^N$, in which $I_{SWI}^i$ denotes the SWI HP-filtered phase images and $I_{QSM}^i$ is the ground truth of QSM images, the loss function can be formulated as:

$$Loss = \frac{1}{2N}\sum_{i=1}^{N}\left\|F(I_{SWI}^i, \theta) - I_{QSM}^i\right\|_2^2 \quad (3)$$

where $F(\cdot)$ and $\theta$ are the mapping and parameters of the CΨ-Net, respectively.

## 3. EXPERIMENTS AND RESULTS

### 3.1. Data Description

Nine healthy volunteers were scanned on a 3T scanner. The mGRE data were recorded with the following parameters: TR/TE = 53/3.63 ms, 8 echoes in total with echo spacing of 5.1 ms, flip angle = 20°, voxel size = 1.1×1.1×1.6 mm$^3$, and scan time = 16 min 18 sec. The raw phase data of mGRE were unwrapped through a Laplacian-based approach [18]. The local phase was then acquired by using the VSHARP algorithm to remove background field [19]. Finally, the labeled QSM were reconstructed by STAR-QSM algorithm. Meanwhile, the SWI data were collected with the following parameters: TR/TE = 28/20 ms, voxel size = 1.1×1.1×1.6, with a scan time of 3 min 38 sec. Nine pairs of SWI phase and QSM images were thus obtained for experiments, all of which have a voxel size of 160×160×88. These images were cropped into patches of 48×48×48 for training the networks.

### 3.2. Experimental Design

To validate the performance of our proposed CΨ-Net for QSM reconstruction from SWI HP-filtered phase, the following algorithms were compared:
1) STAR-QSM [20]: A conventional approach for QSM reconstruction from pre-processed phase data of mGRE;
2) DeepQSM [11]: An improved U-Net-based algorithm that is originally designed for reconstructing QSM from pre-processed local phase;
3) AutoQSM [12]: An improved U-Net-based algorithm that is designed for QSM reconstruction from pre-processed total phase without brain extraction and background phase removal.

We also conducted an ablation experiment to compare CΨ-Net with the following two variants:
1) CΨ-Net w/o DIB: This model replaced each DIB with two standard convolutional layers in CΨ-Net to evaluate the performance of DIB;
2) CΨ-Net w/o MFF: This model removed the resampling operations between branches in CΨ-Net and only preserved the common skip connections to evaluate the effect of the multi-resolution feature fusion scheme.

The leave-one-out cross-validation strategy was used in our experiments. Four different metrics were utilized to evaluate the quality of reconstructed QSM, including peak signal-to-noise ratio (PSNR), structural similarity index (SSIM), root mean squared error (RMSE) and high frequency error norm (HFEN). The results were finally described in the format of mean ± SD (standard deviation). We also calculated the mean susceptibility values within five regions of interest (ROIs) in QSM reconstructed from the testing sample selected for qualitative comparison, which evaluated the accuracy of derived tissue susceptibility. The five ROIs included red nucleus (RN), substantia nigra (SN), globus pallidus (GP), caudate (CAU), putamen (PUT).

### 3.3. Experimental Results

Fig. 2 shows the qualitative results of different reconstruction algorithms from one testing sample for visual comparison. It can be found that the reconstructed susceptibility maps by our proposed CΨ-Net are most close to the ground truth in all views compared with other algorithms. The conventional STAR-QSM fails to reconstruct QSM from the HP-filtered phase data of SWI due to the missing components in phase. The current U-Net-based algorithms, including DeepQSM and AutoQSM, yield image smoothness in reconstructed susceptibility maps, because they cannot learn sufficient information about tissue susceptibility. Even both the variants of CΨ-Net, namely CΨ-Net w/o DIB and w/o MFF, achieve superior performance to DeepQSM and AutoQSM, suggesting the effectiveness of both proposed DIB and multi-resolution feature fusion scheme in CΨ-Net.

We further measured the mean susceptibility values of fives ROIs in the same reconstructed QSM images shown in the qualitative results. It is observed that the results in Fig. 3 are consistent with those in Fig. 2. The proposed CΨ-Net achieves the susceptibility values closest to the ground truth on four ROIs only except Caudate, indicating its superiority. While the conventional STAR-QSM severely underestimates the susceptibility values in all five ROIs.

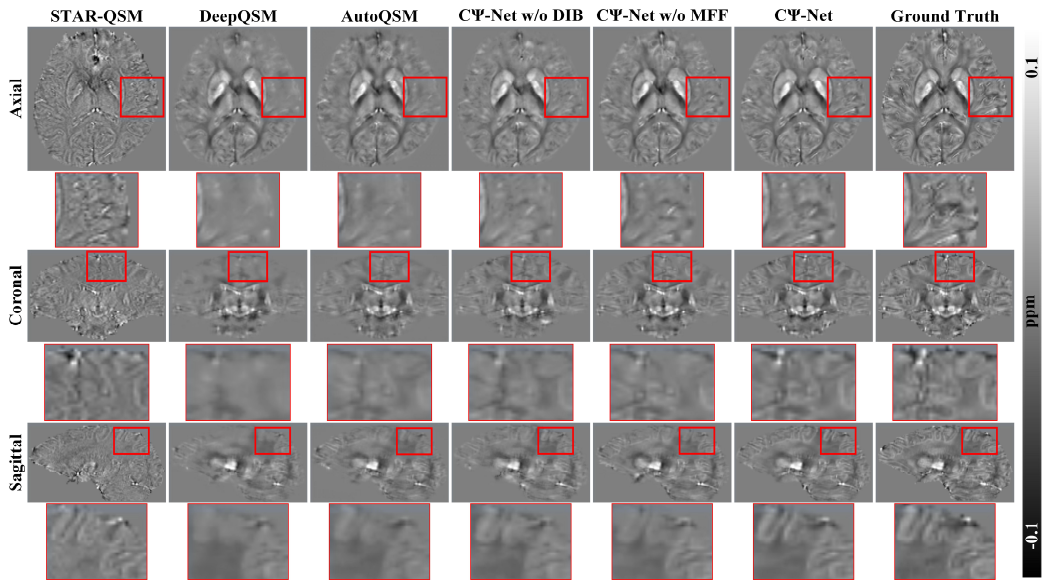

**Fig. 2.** Qualitative results of reconstructed susceptibility maps by different algorithms from one testing sample. The bar denotes the relationship between the grey level and the parts per million (ppm).

Table 1. Quantitative results of different algorithms

| Method | PSNR (dB) | SSIM | RMSE (%) | HFEN (%) |
|---|---|---|---|---|
| STAR-QSM | 28.73 ± 1.25 | 0.8372 ± 0.0505 | 105.61 ± 11.22 | 90.11 ± 8.36 |
| DeepQSM | 31.28 ± 1.09 | 0.8620 ± 0.0393 | 75.51 ± 7.56 | 67.93 ± 7.52 |
| AutoQSM | 31.35 ± 1.16 | 0.8647 ± 0.0403 | 75.00 ± 8.06 | 67.22 ± 7.93 |
| CΨ-Net w/o DIB (ours) | 31.47 ± 1.13 | 0.8672 ± 0.0401 | 74.27 ± 7.84 | 66.28 ± 7.65 |
| CΨ-Net w/o MFF (ours) | 31.49 ± 1.14 | 0.8683 ± 0.0394 | 73.96 ± 8.02 | 65.87 ± 8.02 |
| **CΨ-Net (ours)** | **31.61 ± 1.16** | **0.8703 ± 0.0398** | **73.21 ± 8.35** | **64.79 ± 8.40** |

Table 1 gives the quantitative results of different algorithms. The proposed CΨ-Net outperforms all the compared algorithms with the best mean PSNR of 31.61 dB, SSIM of 0.8703, RMSE of 73.21% and HFEN of 64.79%. Moreover, even both variants of CΨ-Net also achieve superior performance to the other compared algorithms. Specifically, CΨ-Net improves by at least 0.26 dB, 0.0056, 1.79% and 2.43% on PSNR, SSIM, RMSE and HFEN, respectively, comparing with STAR-QSM, DeepQSM and AutoQSM algorithms. CΨ-Net also gets the improvements of at least 0.12 dB, 0.0020, 0.75% and 1.08% on the corresponding indices over the variants in the ablation experiment, which indicates the effectiveness of the proposed DIB and multi-resolution feature fusion scheme。

## 4. CONCLUSION

In summary, we propose a totally new method to directly reconstruct QSM from low-quality SWI HP-filtered phase data, and explores a new value of SWI for quantitative information. A novel CΨ-Net is then developed for this task, in which the newly designed DIBs in the intermediate branch are used to enlarge the receptive fields for learning more susceptibility information from a wider spatial range. A multi-resolution feature fusion scheme implemented by crossed connections is also adopted to capture contextual information necessary for reconstruction. The experimental results indicate the effectiveness of the proposed CΨ-Net, suggesting its potential for clinical applications.

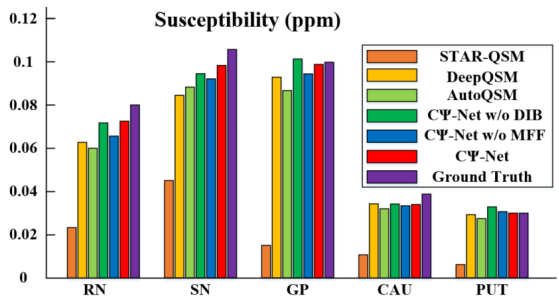

**Fig. 3.** Comparison of different algorithms on the mean susceptibility of five ROIs from one testing sample.

## 5. COMPLIANCE WITH ETHICAL STANDARDS

This study was performed in line with the principles of the Declaration of Helsinki. Approval was granted by the Ethics Committee of Shanghai University (2019).

## 6. ACKNOWLEDGMENTS

This work is supported by the National Natural Science Foundation of China (81871428), the Shanghai Science and Technology Foundation (18010500600), the 111 Project (D20031), the Fundamental Research Funds for the Central Universities (2019QNA5026) and the China Postdoctoral Science Foundation (2020TQ0296). We appreciate the High Performance Computing Center of Shanghai University, and Shanghai Engineering Research Center of Intelligent Computing System (No. 19DZ2252600) for providing the computing resources.